\documentclass[aip,reprint,superscript,floatfix,superscriptaddress,showpacs]{revtex4-1}
\usepackage{amssymb}

\usepackage{epsf}
\usepackage{epsfig}
\usepackage{graphicx}
\usepackage{dcolumn}
\usepackage{braket}
\usepackage{bm}
\usepackage{amsfonts}
\usepackage{amsmath}
\usepackage{amssymb}
\usepackage{color,soul}
\usepackage{wasysym}
\usepackage{mathrsfs}
\usepackage{natbib}
 \usepackage{multirow}
\usepackage{times}

\newcommand{\bea}{\begin{eqnarray}}
\newcommand{\eea}{\end{eqnarray}}

\begin{document}

\title{Valley filtering by a line-defect in graphene: quantum interference and inversion of the filter effect}

\author{L. H. Ingaramo}
\affiliation{Instituto de F\'{\i}sica Enrique Gaviola (CONICET) and FaMAF, Universidad Nacional de C\'ordoba, Argentina}
\author{L. E. F. Foa Torres}
\affiliation{Departamento de F\'{\i}sica, Facultad de Ciencias F\'{\i}sicas y Matem\'aticas, Universidad de Chile, Santiago, Chile}

\begin{abstract}
Valley filters are crucial to any device exploiting the valley degree of freedom. By using an atomistic model, we analyze the mechanism leading to the valley filtering produced by a line-defect in graphene and show how it can be inverted by external means. Thanks to a mode decomposition applied to a tight-binding model we can resolve the different transport channels in $k$-space while keeping a simple but accurate description of the band structure, both close and further away from the Dirac point. This allows the understanding of a destructive interference effect (Fano resonance or antiresonance) on the p-side of the Dirac point leading to a reduced conductance. We show that in the neighborhood of this feature the valley filtering can be reversed by changing the occupations with a gate voltage, the mechanism is explained in terms of a \textit{valley-dependent Fano resonance splitting}. Our results open the door for an enhanced control of valley transport in graphene-based devices.
\end{abstract}
\date{\today}
\maketitle

\section{Introduction} 
Graphene\cite{Geim2007} has two inequivalent Dirac cones related by time-reversal symmetry.\cite{CastroNeto2009,Meunier2016,FoaTorres2014} This endows the electronic states in graphene with a binary, \textit{spin-like}, flavor. Harnessing this valley degree of freedom is an exciting avenue that may lead to new ``valleytronics'' applications. The interest has surged in the last few years not only in the context of graphene \cite{Rycerz2007,Yao2008,Gorbachev2014} but also for other two-dimensional materials like MoS$_2$.\cite{Wu2013}

A very first step for the operation of a valleytronics device is generating a valley polarization, \textit{i.e.} selectively populating a \textit{single} valley. Proposals for achieving such a valley filter include the use of a constriction with zig-zag edges,\cite{Rycerz2007} and scattering against a (8-5-5) line-defect.\cite{Gunlycke2011} The experimental realization of line-defects further contributes to the growing interest in their use as valley filter.\cite{Yazyev2010,Song2012,Rodrigues2012,Yao2015,Xu2014} Indeed, besides their natural occurrence in polycrystalline samples,\cite{Cummings2014,Yazyev2014} different groups have demonstrated controlled growth of 8-5-5 line defects either via chemical vapor deposition on a Nickel step\cite{Lahiri2010} or by Joule heating.\cite{Chen2014}  Filtering via line-defects has been elegantly demonstrated based on symmetry arguments\cite{Gunlycke2011,Gunlycke2014} which were also confirmed through atomistic calculations.\cite{Gunlycke2011,Zhou2016,Liu2013} But previous works were mostly focused on the immediate proximity of the Dirac points where a low energy model can be used.

Here we re-examine valley filtering in graphene with a 8-5-5 line-defect. Our analysis is based on a tight-binding model using a mode-decomposition which helps us to resolve the conductive modes in $k$-space.  This allows us to explore the mechanisms leading to valley filtering both in the neighborhood and also further away from the Dirac point. Close to the Dirac point we recover the results previously reported in the literature. Gating the system to the p-doped side one finds a conductance dip which we interpret as resulting from destructive interference (known in the literature as Fano resonance or antiresonance). Interestingly, we find that around this conductance dip the valley filtering effect can be reversed by applying a gate voltage. This is, if around the Dirac point electrons incident on the defect at a given angle are predominantly transmitted on the $K$ valley, close to the conductance dip one can tune the occupation so that they are transmitted mainly on the $K'$ valley. The mechanism is explained in terms of a \textit{valley-dependent Fano resonance splitting}. This reversal of the filtering effect could be useful, for example, to produce the analog of a spin-valve effect.

In the following we introduce our model Hamiltonian and the scheme used to resolve the scattering in $k$-space produced by the defect.
Later on we examine the operation of the valley filter both close and away from the Dirac point and present our main results.

\section{Hamiltonian model and mode decomposition} 

\begin{figure}[tbph]
\centering
\includegraphics[width=\columnwidth]{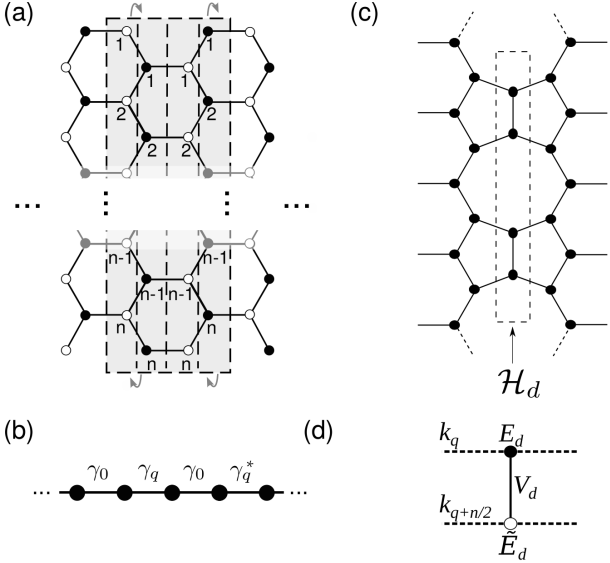}
\caption{(a) Scheme showing a graphene ribbon lattice decomposed into succesive interconnected layers. A mode decomposition of the lattice shown in (a) leads to the independent modes represented in (b). (c) Detail of the line defect considered in the text. The defect couples two of the modes represented in (b) as shown in panel (d).}
\label{fig1}
\end{figure}

Let us consider a simple $\pi$-orbitals Hamiltonian \cite{CastroNeto2009,FoaTorres2014} for graphene:
\begin{equation}
{\cal H}_e=\sum_{i} E_i {\hat c}_i^{\dag} c_i^{}-\sum_{\langle i,j \rangle} \gamma_{i,j} [ {\hat c}_i^{\dag} c_j^{} + h.c. ]
\label{eq-H}
\end{equation}
where ${\hat c}_i^{\dag}$ and ${\hat c}_i^{}$ are the electronic creation and anihilation operators at site $i$, $E_i$ is the site energy and $\langle i,j \rangle$ denote that the summation is restricted to nearest neighbors. The transfer integrals between nearest neighbors is chosen as $\gamma_0=2.7eV$ \cite{FoaTorres2014}. Since we are interested in the two-dimensional limit, we impose periodic boundary conditions along the armchair edge. This makes the quasi-momentum along the vertical direction (see Fig. \ref{fig1}(a)) a good quantum number which we will exploit later on. The defect is modelled by taking into account the changes in the topology of the lattice according to a 8-5-5 linear structure (see Fig. \ref{fig1}(c)) as in Ref. \onlinecite{Gunlycke2011}.

A tight-binding model is, \textit{a priori}, not well suited for obtaining transmission probabilities as a function of the incident angle of the electrons, \textit{i.e.} $k_x$ and $k_y$. Since the Hamiltonian is written in a real space basis, the momentum resolved information is hidden when one follows the standard procedure to compute the transmission probabilities from left to right.\cite{Datta1995,Lewenkopf2013}

To gain resolution in reciprocal space we exploit the periodic boundary conditions and switch to a basis where $k_y$ is well defined, a similar strategy was followed in Refs.~\cite{Zhou2016,Liu2013}. In this basis the Hamiltonian of a pristine armchair ribbon with periodic boundary conditions can be written in block-diagonal form,\cite{FoaTorres2014} where each block can be represented as shown in Fig. \ref{fig1}(b). Since the line-defect doubles the periodicity of the lattice along $y$ it couples only those modes with $k_y$ differing in $|k_y-k_{y'}|=2\pi/2a$, where $a$ is the lattice parameter, which leads to the ladder model represented in Fig. \ref{fig1}(d).

The eigenvectors defining the new basis can be worked out analytically. By choosing slices of the ribbon as shown in \ref{fig1}(a), so that the carbon atoms are along the same vertical line, the ribbon has a periodicity of four of such slices. Then, the problem can be decoupled by changing to the following basis:~\cite{FoaTorres2014} 

\begin{equation}
\label{eq-modes}
	\ket{k_q} = \dfrac{1}{\sqrt{n}} \sum\limits_{j=1}^{n} \exp({\rm i} k_q j a) \ket{j}, 
\end{equation}
where $k_q=2\pi q/na$, $q=1,2,...,n$ ($n$ being an even integer number) and the sum on the right hand side is over the lattice sites on each slice of Fig. \ref{fig1}(a). Therefore, one can see that 
in this basis one gets independent modes as shown in Fig.\ref{fig1}(b). These modes are indexed by $k_q$, the quasimomentum in the vertical direction, and are represented by dimers with hoppings $\gamma_0$ and $\gamma_q=2\gamma_0\exp(-{\rm i}\pi q/n)\cos(q\pi/n)$.

These modes will be mixed by the line defect. The corresponding matrix elements can be obtained by writing the matrix of the defect ${\cal H}_d$ in the basis of Eq. (\ref{eq-modes}).
This gives:

\begin{equation}
\braket{k_q|{\cal H}_d|k_{q'}} = \begin{cases}
    \gamma_0 \cos(k_q a) & \text{if $q=q' \in [1,\ldots,n/2]$}.\\
    -\gamma_0 \cos(k_q a) & \text{if $q=q' \in [n/2+1,\ldots,n]$}.\\
    {\rm i} \gamma_0 \sin(k_q a)  & \text{if $q'=q+n/2$}.\\
    0 & \text{otherwise}
  \end{cases}
\end{equation}
Therefore, as a result of placing the defect in the sample, the dimers are now connected in pairs as represented in Fig.~\ref{fig1}(d). At low energies these two modes are not simultaneusly metallic, which means that an electron cannot be scattered from one valley to the other.

Each leg of the ladder corresponds to a well defined $k_y$. Thus, we still need to resolve the information on $k_x$. This can be done by noticing that the asymptotic states also have a well defined
$k_x$ which is fixed by the dispersion relation of bulk graphene and a boundary condition (direction of incidence). Therefore, an interesting aspect of this representation is that \textit{both} the longitudinal and transversal quasi-momentum of the asymptotic states can be resolved within a tight-binding model.

\section{Valley filtering and inversion of the filter effect} 

\begin{figure*}[t]
\centering
\includegraphics[width=0.95\textwidth]{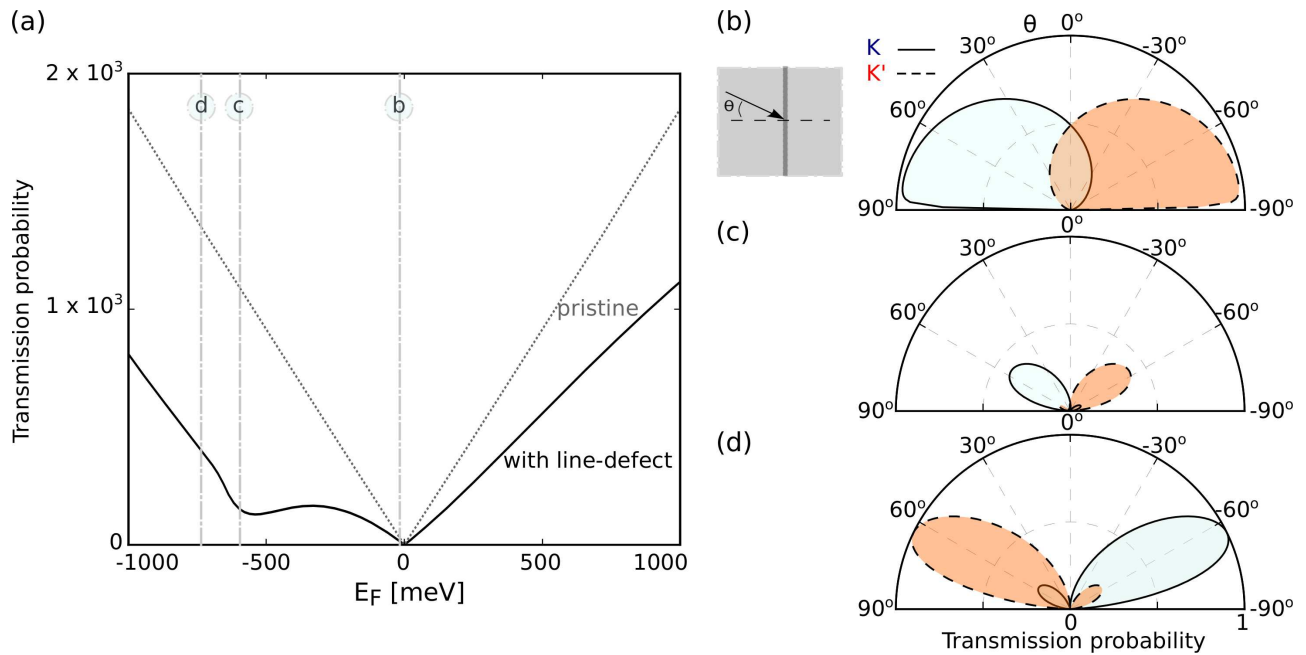}
\caption{(color online) (a) Transmission probability as a function of the incident electronic energy for a ribbon with a line-defect (solid line). The simulations are for a system 93,7 $\mu m$ wide. The transmission for a pristine system of the same dimensions is shown for reference with dotted line. Notice the dip around $E_{f}=-540 meV$. (b-d) Representative polar plots of the transmission probability as a function of the angle of incidence on the line defect ($0^{\circ}$ corresponds to normal incidence, see scheme). (b-d) differ in the energy of the incoming electrons which are marked on panel (a) with grey vertical lines (dash-dot): (b) is for $\varepsilon=-23.3$ meV, (c) is for $\varepsilon=-594$ meV and (d) is for $\varepsilon=-729$ meV. The solid line is for scattering around the $K$ point while the dashed line is for the $K'$ point. (b) and (c) show that the valley polarization gets inverted close to the transmission dip in (a).}
\label{fig2}
\end{figure*}

To motivate our discussion let us examine the transmission probability through a very wide ribbon containing a line defect perpendicular to the transport direction ($x$) as introduced in the previous section. The result of a tight-binding calculation is shown in Fig. \ref{fig2} with a full line, as a reference the result for a pristine system is shown with a the dashed line. Overall, we can see a reduction of the transmission probability consistent with a defect-induced enhanced backscattering. The most prominent difference is the dip observed in the p-doped region (\textit{i.e.} for energies below the Dirac point). We will come back to the physical origin of this feature later on.

Figure \ref{fig2}a provides information on the scattering for different energies of the incident electrons but it does not discriminate between valleys. One might be tempted to infer that the transmission probabilities for electrons entering the sample at a given incident angle is independent on whether the quasimomentum lies close to $K$ or $K'$. Fig. \ref{fig2}b shows that this is not the case. There, we observe the dependence of the transmission probability with the angle of incidence and valley calculated from our tight-binding model at a Fermi energy close to the Dirac point. At normal incidence ($0^\circ$) the value of the transmission is equivalent for electrons from both valleys. In contrast, at high incident angle we see a larger difference of transmission probability between electrons from different valleys, which means a larger valley polarization for those angles. This behavior is stable at energies close to the Fermi level and is consistent with the results reported in Ref. \cite{Gunlycke2011}.

Now, exploiting the capabilities of our atomistic description, we turn to the study of the valley polarization further away from the Dirac point. We are interested, in particular, in the behavior close to the dip observed for negative doping in Fig. \ref{fig2}a. Interestingly, we find that \textit{the valley polarization is reversed around the condutance dip}. Figs. \ref{fig2}c and \ref{fig2}d illustrate the dramatic change in angular dependence of the transmission probability when the Fermi energy is slightly shifted.

Let us examine the changes in the valley filter effect as we move away from the Dirac point. For energies above the Dirac point we find that the lobes shift their dominant angle from grazing angles to angles closer to normal incidence as the energy increases. On the other hand, for negative energies we see a richer behavior: As the energy moves further away from the Dirac point the transmission degrades, especially at normal incidence, the lobes become thinner and their maxima do not reach unity, see Fig. \ref{fig2}-c. Furthermore, for each valley one notices an incipient new lobe in the opposite quadrant, which becomes dominant as the energy is lowered even further, see Fig. \ref{fig2}-d. Therefore, \textit{the operation of the valley filter can be inverted by introducing a small change in the Fermi energy} (\textit{e.g.} through an applied gate voltage).

\begin{figure}[tbph]
\centering
\includegraphics[width=\columnwidth]{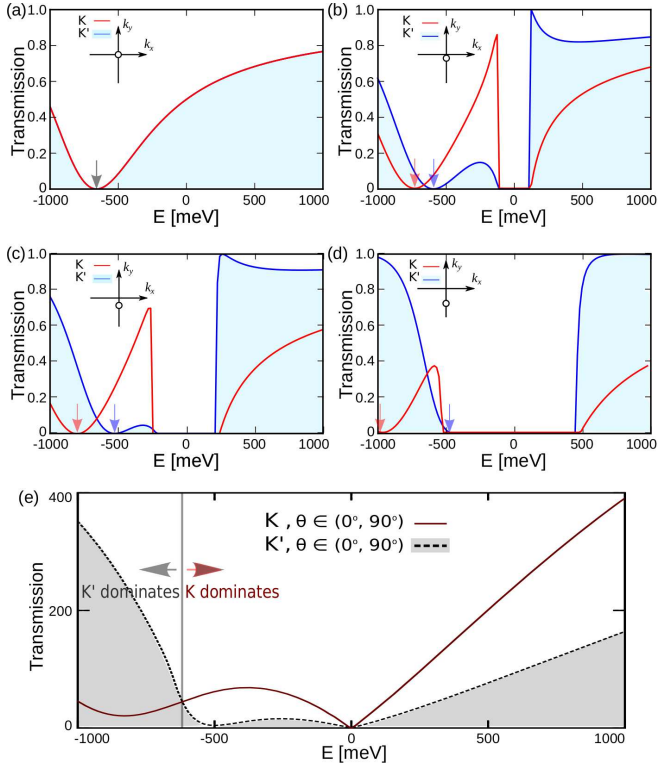}
\caption{(color online) Panels (a-d) show the transmission probabilities for individual channels corresponding to a well defined $k_y$ (represented in the insets, the empty dot being the nearest Dirac point). $k_{y}$ (as measured from the nearest Dirac points) are set to $0$, $2\pi\Delta/a$, $4\pi\Delta/a$, and $6\pi\Delta/a$ respectively ($\Delta=1/130$). e) Total transmission probability for electrons from the $K$ and $K'$ valleys in solid and dashed lines respectively, for the [$0^\circ-90^\circ$] quadrant. Inversion of dominant polarization current occurs at $E_{F} \sim -620meV$ (marked with a vertical grey line).}
\label{fig3}
\end{figure}

The mechanism behind the valley filter inversion turns out to be closely related to that of the conductance suppression. \textit{The conductance dip reported earlier is the manifestation of a destructive interference effect}, known as Fano resonance\cite{Fano1935,Guevara2003,Miroshnichenko2010} or antiresonance\cite{DAmato1989,Levstein1990}. This can be visualized by analyzing the individual modes represented in Fig. \ref{fig1}d, a set of representative transmissions is shown in Fig. \ref{fig3}a-d. The Green's function determining the transmission contains two contributions that can compete with each other: One corresponding to direct transmission from left to right on the same leg of the ladder, and another one which where the mode on the opposite leg (which has a larger energy gap) is explored. This can be captured by a simple two-pole approximation for the retarded Green's function determining the transmission: $G_{R,L} \sim A/(\varepsilon-E_d)+B/(\varepsilon-\tilde{E}_d)$. The pole in the first term is located at $\Re(E_d)>0.5\mid \gamma_0 \mid$ and dominates over the second one which is on the p-doped side. This is because the second pole comes from the non-conducting leg of the ladder and therefore can only provide for a virtual process. The competition between these two poles leads to a destructive interference located closer to the non-dominant pole, on the p-doped side of the spectrum. As shown in Fig. \ref{fig3}, the precise position of this antiresonance or Fano-resonance feature changes slightly from one mode to the other giving the overall behavior shown in Fig. \ref{fig2}a.

Now, the question is how is this destructive interference linked to the inversion of the valley filter effect. To rationalize it let us examine more closely the panels in Fig. \ref{fig3}. One can see that while moving away from the Dirac point in a chosen direction (as in Figs. \ref{fig3}b-d) the poles in the two-pole approximation get closer together in one valley while they get further apart in the other. \textit{Therefore, as one moves away from the Dirac point, the destructive interferences on each valley, which are degenerate in Fig.\ref{fig3}a, follow the movement of the pole on the p-doped side (\ref{fig3}b-d), thereby suffering a valley-dependent splitting} (the position of the Fano resonances is marked with arrows in Fig.\ref{fig3}). This splitting, in turn, produces the observed change in the valley polarization direction around the destructive interference. This behavior can also be verified analytically from the ladder model.

To better visualize the valley-dependent splitting of the Fano resonance let us consider the sum of all the transmission probabilities from electrons with incident angles in one quadrant ($0^{\circ} < \theta < 90^{\circ}$) on each valley separately. This is shown in Fig. \ref{fig3}e where the total transmission for the $K$ and $K'$ valleys are plotted in solid and dashed lines respectively. Near the charge neutrality point, transmission from $K$ valley is dominant. This is consistent with the results for low energy observed in Fig. \ref{fig3}a-d. At $E_{F} \sim -620meV$ (gray vertical line) and one has balanced transmission from both valleys, which means zero polarization for the whole quadrant. In contrast, for  $E_{F} < -620meV$ the valley polarization is inverted. 

\section{Conclusions} 

We have shown a new mechanism leading to an inversion of the valley filtering effect in graphene: the \textit{valley-dependent splitting of the Fano resonances}. Our results for graphene with a line-defect show that this could be achieved on the p-doped side where a destructive interference is present. This would allow for better control of valley filtering in graphene and one could envisage, for example, the realization of the analog of a spin-valve\cite{Wakabayashi2002} by using two valley filters in series with their easy valley axis inverted.

Indeed, previous works showed that line defects in series could provide for an enhanced control of the valley polarization.\cite{Liu2013} Nonetheless, since the authors focused in the vicinity of the Dirac point, where the mechanism that we propose is not active, the use of creative configurations together with a gate voltage remains as an interesting problem for further study.

We acknowledge partial funding by Program 'Inserci\'on' 2016, University of Chile and support from CONICET (Argentina). LEFFT acknowledges the support of the Abdus Salam ICTP associateship program. LEFFT is on leave from Universidad Nacional de C\'ordoba (Argentina) and CONICET.
\noindent



%

\end{document}